\magnification=1200
\baselineskip18pt

\def\a{\alpha}\def\e{\epsilon}
\def\f{\phi}
\def\l{\lambda}\def\o{\omega}\def\r{\rho}
\def\y{\eta}

\def\L{\Lambda}
\def\O{\Omega}

\def\na{\nabla}
\def\mo{{-1}}\def\ha{{1\over 2}}
\def\qu{{1\over 4}}

\def\const{{\rm const}}

\def\mn{{\mu\nu}}  
\def\ds{ds^2=}\def\sg{\sqrt{-g}}

\def\st{spacetime }
\def\fe{field equations }\def\bh{black hole }
                                   
\def\bg{background }\def\gs{ground state }

\def\ct{conformal transformation }
\def\ns{naked singularity }
\def\rep{representation }
\def\sch{Schwarzschild }\def\min{Minkowski }\def\ads{anti-de Sitter }
\def\poi{Poincar\'e }
\def\KK{Kaluza-Klein }\def\des{de Sitter }
 
\def\GR{general relativity }\def\GB{Gauss-Bonnet }\def\CS{Chern-Simons }
\def\EH{Einstein-Hilbert }

\def\section#1{\bigskip\noindent{\bf#1}\smallskip}
\def\nota{\footnote{$^\dagger$}}

\def\PL#1{Phys.\ Lett.\ {\bf#1}} 
\def\PRL#1{Phys.\ Rev.\ Lett.\ {\bf#1}} 
\def\PR#1{Phys.\ Rev.\ {\bf#1}} 
\def\NP#1{Nucl.\ Phys.\ {\bf#1}} 
\def\JMP#1{J.\ Math.\ Phys.\ {\bf#1}}

\def\ref#1{\medskip\everypar={\hangindent 2\parindent}#1}
\def\beginref{\begingroup
\bigskip
\centerline{\bf References}
\nobreak\noindent}
\def\endref{\par\endgroup}

\def\mm{{mn}}\def\hmn{H_\mm}\def\ab{{ab}}\def\mt{{-2}}
\def\Lh{{\L\over2}}\def\Lq{{\L\over4}}
\def\ff{{f'\over f}}\def\hh{{h'\over h}}
\def\cR{{\cal R}}\def\cT{{\cal T}}\def\cL{{\cal L}}
\def\uno{{(1)}}\def\due{{(2)}}\def\zero{{(0)}}


{\nopagenumbers
\line{September 1996\hfil INFNCA-TH9619}
\vskip40pt
\centerline{\bf Black hole solutions in four-dimensional topological gravity}
\vskip40pt
\centerline{{\bf S. Mignemi}\nota{e-mail:MIGNEMI@CA.INFN.IT}}
\vskip10pt
\centerline {Dipartimento di Matematica, Universit\`a di Cagliari}
\centerline{viale Merello 92, 09123 Cagliari, Italy}

\vskip40pt
{\noindent 
We study spherically symmetric solutions of a four-dimensional theory of 
gravity with a topological action, which was constructed as a Yang-Mills
theory of the Poincar\'e group and can be considered a generalization to
higher dimensions of well-known two-dimensional models. We also discuss
the perturbative degrees of freedom and the properties of the theory under
conformal transformations.}
\vfil\eject}

\section {1. Introduction}
In recent years, there has been a great interest in the possibility of writing
down the theory of gravity as a Yang-Mills theory of the \poi or (anti)-\des 
group, motivated by the fact that in this form the theory resembles the other 
known fundamental interactions and should be renormalizable [1,2]. 
This goal has
been achieved in two and three dimensions by using actions of topological 
origin. In the three-dimensional case, one makes use of a \CS type action [1],
while in two dimensions, in order to construct an action, one is forced to 
introduce a multiplet of auxiliary scalar fields besides the gravitational
variables [2]. After a suitable identification of the Yang-Mills fields with
the geometric quantities of a riemannian manifold, these actions result to be
linear in the curvature tensor of the manifold, and give rise to models which
are the closest possible generalization of \GR to two or three dimensions.

Some time ago, Chamseddine [3] showed that this approach can be generalized 
to higher dimensions at the cost of introducing actions containing higher
powers of the Riemann tensor. 
In particular, for odd dimensional theories, one generalizes the \CS action
by using higher dimensional \CS invariants, while in even dimensions, 
where a \CS action cannot be defined,
one makes recourse to \GB invariants coupled to auxiliary scalar fields.

These models are expected to be renormalizable, since no potential 
counterterms exist [3]. Moreover, as in the lower-dimensional cases,
different phases are expected to arise,
depending on the vacuum expectaction values of the fields. In particular, a
topological phase with vanishing expectation value for the 
vielbeins, and some broken phases including the ordinary spacetime of general
relativity should exist.

In this paper, we shall be interested in the four-dimensional case, which,
if one excludes possible mechanisms of spontaneous compactification, is the
physically relevant one. We shall limit ourselves to the study of the
"riemannian" phase, corresponding to vanishing torsion and nonvanishing
expectation value for the vierbeins.
The physics of the four-dimensional models will turn out to be quite different 
from that of Einstein gravity. 
In particular, the field equations do not reduce to those of \GR in any
limit. Moreover, one of the properties of the lower-dimensional models which 
seems to
be valid also in higher dimensions is the fact that these theories
do not possess any propagating degrees of
freedom around maximally symmetric backgrounds.

In order to investigate more deeply the physical implications of these
models, we shall study the spherically symmetric solutions of the
field equations in the riemannian sector of the theory. We show that in
general the \fe do not determine completely the form of the solutions.
We also examine the propagation of small excitations around the relevant
backgrounds and discuss the behaviour of the model under conformal 
transformations.

\section{2.1 Gauge theories of gravity in two dimensions}
Before describing the 4-dimensional theory, we briefly review the properties
of the 2-dimensional models [2], which have many similar features.
In two dimensions, the gauge invariant action can be written
$$I_2=\int_{M_2}\e_{ABC}\ \y^AF^{BC}\eqno(2.1)$$
where $\e_{ABC}$ is a totally anntisymmetric tensor,
$F^{AB}$ is the field strength of $SO(1,2)$ or $ISO(1,1)$ and $\y$ is
a triplet of scalars in the fundamental \rep of the gauge group. The group
indices run from 0 to 2.

The action is anomaly free and finite [2]. Connection with gravity is made
through the identification $A^{ab}=\o^{ab}$, $A^{a2}=e^a$, where $\o^{ab}$ and 
$e^a$ are the spin connection and vielbeins of the two-dimensional manifold
($a,b=0,1$).
This yields $F^{ab}=\cR^{ab}+\l e^a e^b$, $F^{a4}=\cT^a$, where $\cR^{ab}$
and $\cT^a$ are the curvature and the torsion 2-forms on $M_2$ and $\l= 0, 
\pm1$ for the \poi or (anti-)\des group respectively.

With these identifications, the action can then be written 
$$\int_{M_2}\e_{ab}(\y^2\cR^{ab}+\y^a\cT^b+\l\y^2e^ae^b)\eqno(2.2)$$
The field equations obtained varying with respect to $\y^a$ imply the 
vanishing of the torsion. One can then define as usual the
riemannian connection and the metric $g_\mm=\y_{ab}e^a_me^b_n$, in terms of
which the action becomes
$$I_2=\int_{M_2}\sg\ \y[R(g_\mm)+\l]\eqno(2.3)$$
where $R$ is the Ricci scalar and $\y\equiv\y^2$.
In dimensions other than two the scalar field $\y$ in front of $R$ can be
reabsorbed by rescaling the metric $g_\mm$. This however is not possible in
two dimensions.
This is obvious in view of the fact that $\sg R$ is a total derivative in 
two dimensions and is in fact proportionalto the Euler form. 
A conformal transformation of the metric by a function of the scalar has 
essentially the effect of adding some $\y$-derivative terms to the action.
For $g_\mn\to\y g_\mn$, for example, the action transforms into
$$I_2=\int_{M_2}\sg\ \y[R+4{(\na\y)^2\over\y^2}+\l\y]\eqno(2.4)$$

A weak field expansion of the action (2.3) shows that it possesses no 
propagating degrees of freedom. Nevertheless, it admits \bh solutions which,
if $\l\ne0$, in the \sch gauge read
$$\ds-\left({\l x^2\over2}-c\right)dt^2+\left({\l x^2\over2}-c\right)^\mo 
dx^2\qquad\qquad\y=\l x\eqno(2.5)$$
The metric is that of \des or \ads \st depending on the sign of $\l$.
The integration constant $c$ is proportional to the ADM mass of the solution.
In particular, in the \ads
case, if $c>0$ an event horizon is present at $x=\sqrt{2c/\l}$ and the 
solution can be interpreted as a regular black hole [4].

If $\l=0$, instead, the spacetime is flat and
$$\ds-(ax+b)dt^2+(ax+b)^\mo dx^2\qquad\qquad\y=cx+d\eqno(2.6)$$
with $a$, $b$, $c$ and $d$ integration constants. Also these solutions can
be interpreted as black holes [4].

\section{2.2 Gauge theories of gravity in 4 dimensions}
In four dimensions, the relevant groups are the \poi group $ISO(1,3)$, the
\des group $SO(1,4)$ and the \ads group $SO(2,3)$.
In order to construct the action, one can make use of the
totally antisymmetric group invariant tensor
$\e_{ABCDE}$, in the case of \des or \ads groups, while the \poi case is
recovered by In\"onu-Wigner contraction. The group indices run from 0 to 4.
The action is given by [3]:
$$I=\int_{M_4}\e_{ABCDE}\ \y^AF^{BC}F^{DE}\eqno(2.7)$$
where $\y^A$ is a multiplet of scalars in the fundamental 
\rep of the gauge group and the field strength 2-form $F^{AB}$ is defined as
$F^{AB}=dA^{AB}+A^{AC}A^{CB}$, the 1-form $A^{AB}$ being the connection of
the gauge group.

In analogy with the two-dimensional case, one can then identify 
$A^{ab}=\o^{ab}$, $A^{a4}=e^a$, where $\o^{ab}$ and 
$e^a$ are the spin connection and  vielbeins of the four-dimensional manifold.
This implies that $F^{ab}=\cR^{ab}+\l e^a e^b$, $F^{a4}=\cT^a$, where 
$\cR^{ab}$
and $\cT^a$ are the curvature and the torsion 2-forms of the 4-dimensional
manifold and $\l=1$ for $SO(2,3)$,
$\l=-1$ for $SO(1,4)$ or $\l=0$ for $ISO(1,3)$.
The action can be written in terms of the geometrical quantities as
$$I=\int_{M_4}\e_{abcd}\left[\y^4(\cR^{ab}+\l e^ae^b)(\cR^{cd}+\l e^ce^d)+
4\y^a\cT^b\cR^{cd}\right]\eqno(2.8)$$
The field equations obtained by varying (2.8) with respect to the $\y$ fields 
are thus
$$\eqalign{&\e_{abcd}\ \y^4(\cR^{ab}+\l e^ae^b)(\cR^{cd}+\l e^ae^b)=0\cr
&\e_{abcd}\cT^b(\cR^{cd}+\l e^ce^d)=0\cr}\eqno(2.9)$$
It is easy to see that the second equation is satisfied if the torsion 
vanishes. 
Contrary to the 2-dimensional case, however, this is not the unique solution.
Nevertheless, in the following we shall limit our attention only to this 
sector.

When the torsion vanishes, the connection can be expressed as usual in terms
of the vielbeins, and then one can write the action in the metric formalism as
$$I=\int_{M_4}\sqrt{-g}\ d^4x\ \y(S+4\l R+24\l^2)\eqno(2.10)$$
where $\y$ is the scalar field $\y^4$, $R$ is the Ricci scalar on $M_4$, and
$S$ is the \GB term
$$S=R_{mnpq}R_{mnpq}-4R_{mn}R_{mn}+R^2\eqno(2.11)$$
which is proportional to the Euler density in four dimensions. $\sg S$ is 
therefore a total derivative in 4 dimensions and so can appear in the action
only coupled to a scalar. This term has already been introduced in several 
contexts, due to its peculiar properties. In particular, it is the only
four-dimensional
higher-derivative term that does not involve derivatives higher than 
second order in the field equations [5]. Moreover it does not introduce new 
degrees of freedom into the action, besides the graviton [6]. For this reasons,
it has been considered in \KK models [7]
and appears in low-energy string effective actions [8,9].

\section{3.1. The Poincar\'e group}
When $\l=0$, only the first term in (2.10) survives. The action is therefore a
total derivative times a scalar and does not include terms proportional to 
the \EH action. In the following, in analogy to some 
2-dimensional models, we shall also consider
the slightly more general case in which a cosmological constant term $\L\y$
is added to the action: 
$$I=\int\sqrt{-g}\ d^4x\ \y(S+\L)\eqno(3.1)$$
Of course, if $\L\ne 0$, the action is not gauge-invariant.
The field equations arising from the action (3.1) are:
$$\hmn=\Lh\y g_\mm\eqno(3.2)$$
$$S=\L\eqno(3.3)$$
where
$$\eqalign{\hmn=&2[4R_{p(m}\na_{n)}\na_p\y-2R_\mm\na^2\y-2g_\mm R_{pq}
\na_p\na_q\y\cr
&-4R\na_{(m}\na_{n)}\y+g_\mm R\na^2\y+2R_{qmpn}\na_p\na_q\y]\cr}
\eqno(3.4)$$
If we denote by $H$ the trace of $\hmn$:
$$H=2(2R_{pq}\na_p\na_q\y-R\na^2\y)=4G_{pq}\na_p\na_q\y\eqno(3.5)$$
the trace of (3.2) gives:
$$H=\L\y\eqno(3.6)$$
The tensor $\hmn$ satisfies the equivalent of a Bianchi identity:
$$\na_m\hmn=\ha\na_n H$$
From (3.6) then follows that
$$\na_m\hmn=\Lh\na_n\y\eqno(3.7)$$
consistently with (3.2).

In the following we shall be mainly interested in spherically symmetric
solutions of the field equations. If one assumes for the metric the static
\sch form 
$$\ds-f^2(r)dt^2+h^{-2}(r)dr^2+r^2d\O^2\eqno(3.8)$$ 
the \fe (2) yield:
$$\eqalign{&(1-h^2)\y''+\left(\hh-3hh'\right)\y'=-\Lq r^2\y\cr
&(1-3h^2)\ff\y'=-\Lq r^2\y\cr
&h^2f'\y''+\left(\hh f'+h^2f''\right)\y'=\Lq r\y\cr}\eqno(3.9)$$
Due to the identity (3.7), only two of these equations are independent.
Moreover,
recalling that $\sqrt{-g}S$ is a total derivative, one can write (3.3) as:
$$8[f'h(1-h^2)]'=-\L fh^\mo r^2\eqno(3.10)$$

\section{3.2 $\bf\L=0$}
In the case of vanishing cosmological constant $\L$, a trivial solution of the
\fe is given by flat space. In this case $\y$ remains undetermined, since 
$\hmn$ vanishes independently of $\y$.
The possibility of choosing freely one arbitrary function in the solution 
seems to be a general feature of the models we are considering. All solutions 
we were
able to find depend on one arbitrary function and only by imposing futher
constraints can one obtain solutions in closed form.

The flat space solution with constant $\y$
can be considered as the ground state of the theory. 
Let us now look for more general solutions of the field equations.
When $\L=0$, one has from (3.10)
$$f'h(1-h^2)=\const=\a\eqno(3.11)$$
For $\a=0$, two possible solutions emerge: either $h=1$ or $f$ is a constant,
which can be put to 1 by rescaling $t$.

Let us consider the first case: the equations (3.9) are then
satisfied if $f'\y'=0$.
Hence one recovers the flat solution if $f'=0$, or a solution with arbitrary
$f$ and constant $\y$. The second case corresponds to a spacetime with flat
spatial sections, and metric of the form
$$\ds-f^2(r)dt^2+dr^2+r^2d\O^2\eqno(3.12)$$
The arbitrarity of $f$ does not consent in this case 
to draw conclusions about the causal 
structure of the solutions. Also the newtonian limit is undetermined, since
the newtonian potential is related to $f$.

In the case $f=1$, the equations (3.9) reduce to
$$(1-h^2)\y''+\left(\hh-3hh'\right)\y'=0\eqno(3.13)$$
This equation admits solutions with arbitrary $h$ and $\y$ either constant
or given by
$$\y'={C\over h(1-h^2)}\eqno(3.14)$$
with $C$ an integration constant.
The metric has in this case the form
$$\ds-dt^2+{dr^2\over h^2(r)}+r^2d\O^2\eqno(3.15)$$
and is therefore the direct product of the time coordinate with an arbitrary
spherically symmetric 3-space. Again, the properties of the specific 
solutions depend on the form of the function $h$. However, from the geodesic
equation is easy to see
that a point particle does not experience any gravitational force in this 
metric.

We pass now to consider the case in which $\a\ne 0$. In this case one cannot
find the general solution of the field equations.
Some special solutions can nevertheless be obtained by imposing suitable
ans\"atze.
For example, a solution is easily obtained by putting $h^2=1/3$, in order to
satisfy the second of equations (3.9). The other \fe then imply that $f=ar$
and $\y=br+c$, with $a$, $b$, $c$, integration constants. This solution is
regular everywhere, except for a conical singularity at the origin.

Another class of solutions is obtained for $\y=\const$. In this case,
equations (3.9) are trivially satisfied and from (3.11) one has
$$f'={\a\over h(1-h^2)}\eqno(3.16)$$
with arbitrary $h$.
A special solution can be obtained if one requires $f$=$h$, as for the \sch
metric of \GR. In this case, (3.11) can be written as
$$[(1-h^2)^2]'=a^2\eqno(3.17)$$
which can then be easily integrated yielding
$$h^2=f^2=1\pm a\sqrt{r-b}\eqno(3.18)$$
with $a$, $b$ integration constant. 
The curvature invariants built from this metric are singular at $r=0$ and
$r=b$. This indicates the presence of a physical singularity at these points. 
The singularity at $r=b+a^\mt$ arising when the minus sign is chosen in (3.18),
is instead simply a coordinate singularity, corresponding to a horizon.
It is easy to check that in any case, however, the singularities at $r=0$
or $r=b$ are naked, so that no regular \bh solution exists of this form.
It may be interesting to notice, however, that this solution corresponds 
to a gravitational force decreasing as $r^{-1/2}$ at infinity in
the weak field limit.

\section{3.3 $\bf\L\ne0$}
In the case $\L\ne0$, maximally symmetric solutions exist if $\L>0$ and are 
given by de Sitter
or anti-de Sitter space with curvature $\pm\sqrt\L/24$. In this case, 
the scalar field
$\y$ must vanish according to the field equations. This solution can be
considered the \gs of the theory.

More general solutions are much more difficult to obtain than in the $\L=0$
case, since now (3.10) cannot be integrated explicitly.
One can however obtain a first integral of (3.10) by writing the metric as:
$$\ds-Adt^2+{dr^2\over AR^4}+R^2d\O^2\eqno(3.19)$$
with $A=A(r)$, $R=R(r)$. In this case,
$$A'R^2(1-AR^4R'^2)=2\L r+a\eqno(3.20)$$
and for any given $R$ one can obtain $A$ by solving a non-linear first order
differential equation. In general it is not possible to give the solution
in a closed form. A special (unphysical) case is given by
$R=r^{-1/4}$, $A=\L r^{5/2}$,
which in different coordinates can be written as
$$\ds-\L\r^{5/3}dt^2+(\L\r^{5/3})^\mo d\r^2+\r^{-3}d\O^2\eqno(3.21)$$

An explicit solution can also be obtained, as in the $\L=0$ case, by
imposing the ansatz $f=h$. One has now from (3.10):
$$[(1-h^2)^2]'={\L r^3\over6}+a\eqno(3.22)$$
and then
$$h^2=f^2=1\pm\sqrt{{\L r^4\over24}+ar+b}\eqno(3.23)$$
with $a$ and $b$ integration constants. With this form of the metric functions,
the \fe (3.9) can be satisfied only if $\y=0$.
If $a=b=0$, the metric reduces to that of \ads or \des spacetime, depending
on the sign in (3.23). Otherwise,
the solution is \des or \ads only asymptotically.
The curvature invariants are
singular at $r=0$ and at the zeroes of ${\L r^4\over24}+ar+b$. In general
also some horizons may be present, depending on the signs of the parameters
of the solutions. However, the only case in which no \ns arise is
when the minus sign is taken in (3.23) and $\L>0$, $a<0$, $b>\max
\{1,\ 2\L^{-1/3}(3a/4)^{4/3}\}$. In this case
the metric is qualitatively similar to that of a \bh in \des spacetime.

\section{4. The (anti)-de Sitter group}
In the (anti)-de Sitter case, the action takes the form:
$$I=\int\sqrt{-g}\ d^4x\ \y(S+4\l R+24\l^2)\eqno(4.1)$$
and the field equations are 
$$\hmn+4\l\y G_\mm+4\l K_\mm-12\l^2\y g_\mm=0\eqno(4.2)$$
$$S+4\l R+24\l^2=0\eqno(4.3)$$
where $G_\mm$ is the Einstein tensor and
$$K_\mm=-(\na_m\na_n-g_\mm\na^2)\y\eqno(4.4)$$
Even if a \EH term is now present in the action, the theory is quite different 
from general relativity, since the coupling with the scalar field $\y$ yields 
the constraint (4.3), which has no counterpart in the Einstein theory.                           

It is difficult to find a general spherically symmetric solution of the 
field equations.
However, from (2.9) one sees that the field equations can be satisfied if 
$\cR^{23}=-\l e^2e^3$. Substituting the ansatz (3.8) one can then check that 
any metric of the form
$$\ds-f(r)dt^2+{dr^2\over 1+\l r^2}+r^2d\O^2\eqno(4.5)$$
with arbitrary $f$ is a solution of the field equations. 
These solutions are therefore a direct generalization of the solutions (3.12), 
which correspond to
the $\l=0$ limit. For $\l<0$, they describe a spacetime whose spatial sections 
are 3-spheres. In particular, in the special case $f(r)=h(r)=1+\l r^2$, one
obtains the \des solution. In the general case, instead, the 
properties of the solutions depend on the specific form of the function $f$.
Analogous considerations hold for $\l>0$.

For the solutions (4.5), the scalar field equations (4.2) become, if $f\ne h$,
$${\y'\over r}=\l\y\qquad\qquad\y''+{\l r\y'\over 1+\l r^2}=\l\y\eqno(4.6)$$
The only solution of this system is $\y=0$. If $f=h$, instead, all equations 
are identically satisfied and $\y$ can be an arbitrary function of $r$.

We point out that consideration of the ansatz $\cR^{01}=-\l e^0e^1$ does not
generate any new solutions.

\section{5. Conformal transformations}
In this section, we briefly discuss the effect of a conformal transformation
on the action (2.10). In four dimensions, a suitable \ct allows one to remove the
scalar field in front of the \EH term, but the effect on the \GB term is
simply to add some term containing fourth order derivatives of the scalar
fields to the action. This is analogous to what happens in two dimensions
with the \EH term, as discussed in section 2.1.
The transformation of course affects the geometrical properties of the theory, 
but not
the field theoretical ones, since it simply amounts to a redefinition of the
fields. In particular, the field equations will contain at most second order
derivatives of the fields, even if not linearly.

In more detail, if we consider the action (2.10) and perform the \ct $g_\mm\to
e^{2\f}g_\mm$, where $\y=e^{2\f}$, in order to eliminate the scalar field in 
front of $R$, the action becomes:
$$\eqalign{I=4\l\int&\sg\ d^4x\big\{R+6(\na\f)^2+6\l e^{2\f}\cr+&{1\over4\l}
e^{-2\f}[S+2R(\na\f)^2-2R_{ab}\na_a\f\na_b\f-8(\na^2\f)^2+32\na^2\f(\na\f)^2]
\big\}\cr}\eqno(5.1)$$
Hence, in the $\l\ne0$ case, the rescaled action assumes the usual \EH form
with some non-minimal corrections and displays some similarity with the 
effective four-dimensional string lagrangians [9]. One might speculate
that effective string lagrangians can be obtained in this way from a suitable
modification of the gravitational gauge group, as in two dimensions [10].
When $\l=0$, instead, only the terms in square brackets survive and the action 
changes only by higher derivative terms in the scalars. It may be interesting 
to notice that the peculiar combination of these scalar terms does not lead to
derivatives higher than second in the field equations.

In the \poi case, solutions of (5.1) can be easily obtained from those of the
conformally equivalent action by simply applying the \ct to the metric. 
In the (anti-)\des case, instead, this is not possible, since we were
only able to find solutions with vanishing scalar field, for which the \ct
becomes singular.

\section{6. Linearization}
In order to investigate the particle content of the models discussed above,
it is interesting to consider the propagation of excitations around given 
backgrounds. It is well known that the results depend on the specific \gs 
around which one expands: for example, in the unbroken phase of the full 
theory, $e=\o=0$, no propagation can arise. Here we consider the theory in 
its metric form (2.10). As we shall see, also in this case no propagating
degree of freedom is present, at least for the most natural choice of the \gs.
This is similar to the two-dimensional case, where it is well-known that 
no propagation takes place.

In order to investigate this point, one must expand the metric and the 
scalar around a fixed ground state, $g^\zero_\mm$, $\y^\zero$, as
$$g_\mm=g^\zero_\mm+h_\mm,\qquad\qquad\y=\y^\zero+\f$$
For generic \bg the linearized action has a complicated form. However, for
maximally symmetric backgrounds, one has in four dimensions (the superscripts 
denote the order of the terms in the expansion):
$$\eqalign{(\sg R)^\uno=&-(\na^2h_{aa}-\na_a\na_b h_\ab)-3\l h_{aa}\cr
(\sg S)^\uno=&\ 4\l(\na^2h_{aa}-\na_a\na_b h_\ab)\cr
(\sg R)^\due=&-\qu(h_{ab}\na^2h_{ab}-h_{aa}\na^2h_{bb}+2h_{cc}\na_a\na_b
h_{ab}-2h_{ac}\na_a\na_bh_{bc})\cr &-\l(2h_\ab h_\ab-\l h_{aa}h_{bb})\cr
(\sg S)^\due=&\ 0\cr}$$
where $(\sg S)^\due$ vanishes in four dimensions because of the topological 
properties of the \GB term [6].

For $\l=0$, the most natural choice for the \bg is given by the flat \min
metric with constant scalar. It follows that $(\sg\y S)^\due\sim(\sg S)^\uno
\y^\uno=0$, since $(\sg S)^\uno=0$ if $\l=0$.
Analogously, for $\l\ne0$, defining $\cL=S+4\l R+24\l^2$,
$$\eqalign{\sg\cL)^\uno=&\ 0\cr
(\sg\cL)^\due=&-\qu(h_{ab}\na^2h_{ab}-h_{aa}\na^2h_{bb}+2h_{cc}\na_a\na_b
h_{ab}-2h_{ac}\na_a\na_bh_{bc})\cr}$$
From the results of section 4, the most natural choice 
for the \gs appears to be $\y^\zero=0$. Hence
$$(\sg\y\cL)^\due=\y^\uno(\sg\cL)^\uno+\y^\zero(\sg\cL)^\due=0$$
Therefore, for any value of $\l$ no propagating degree of freedom is present in the 
spectrum of the theory, since the term of the lagrangian quadratic in the 
perturbations vanishes. However, if one had expanded around $\y$= const in
the \des case, a graviton would have appeared in the spectrum, since in that 
case $(\sg\y\cL)^\due\sim(\sg\cL)^\due$ which is proportional to the graviton
propagator. 

Also interesting is the case of the non-gauge-invariant action (3.1) with 
$\L\ne0$.
In this case, taking an (anti-)\des\bg with $\y^\zero=0$, 
$$(\sg\y(S+\L))^\due=\y^\uno(\sg(S+\L))^\uno=\f(\na^2 g_\ab-\na_a\na_b)h_\ab
+\Lh\f h_{aa}$$
In the gauge $\na_bh_\ab=\ha\na_bh_{aa}$, the propagator reduces to $\ha\f
(\na^2+\L)h$, with $h=h_{aa}$, or diagonalizing
$$\ha[(\f-h)(\na^2+\L)(\f-h)]-\ha[(\f+h)(\na^2+\L)(\f+h)]$$
This corresponds to two propagating massive scalar degrees of freedom, one of 
which is a ghost.
\section{7. Final remarks}
We have discussed some properties of a Yang-Mills theory of gravity in four 
dimensions. From the results obtained, it appears that the \fe equations for 
the metric are somehow underdetermined, at least in the case of
a spherically symmetric ansatz.
Only one of the metric fields is in fact determined by the field equations.
On the contrary, it is quite difficult to find non-trivial solutions 
for the scalar field. In particular, in the \des and \ads cases, all the 
solutions we have found imply vanishing scalar field. A consequence of this
fact is that no propagating degrees of freedom are present in the spectrum of 
the theory. This was perhaps to be expected in view of the topological nature
of the action of the theory.

Of course, as in all gravitational theories, it is essential for the physical
interpretation to consider the coupling to matter. If one assumes that the
scalar field $\y$ is physical, and not simply a Lagrange multiplier,
the coupling is not uniquely determined.
This is a common feature in gravity-scalar theories.
However, even introducing non-trivial couplings of the matter with the scalar
field, it does not seem possible to obtain a newtonian limit in the framework
of these models. A more physical behaviour may perhaps be obtained by
considering conformally related models, as those discussed in section 4,
in which an Einstein-Hilbert term is present in the action for $\l\ne0$.

Some of the problems discussed so far might be solved by considering the
more general sector containing non-trivial torsion fields, which could give 
rise to more general solutions of the field equations.

\beginref
\ref [1] E. Witten, \NP{B311}, 46 (1989);
\ref [2] T. Fukuyama and K. Kamimura, \PL{B160}, 259 (1985);
K. Isler and C. Trugenberger, \PRL{63}, 834 (1989);
A.H. Chamseddine and D. Wyler, \PL{B228}, 75 (1989);
\ref [3] A.H. Chamseddine, \NP{B346}, 213 (1990);
\ref [4] M. Cadoni and S. Mignemi, \PR{D51}, 4319 (1995); \PL{B358}, 217 
(1995);
\ref [5] D. Lovelock, \JMP{12}, 498 (1971);
\ref [6] D.G. Boulware and S. Deser, \PRL{55}, 2656 (1985);
\ref [7] J. Madore, \PL{A110}, 289 (1985);
F. M\"uller-Hoissen, \PL{B163}, 106 (1985);
\ref [8] B. Zwiebach, \PL{B156}, 315 (1985);
\ref [9] D.J. Gross and J.H. Sloan, \NP{B291}, 41 (1987);
\ref [10] H. Verlinde, in {\it Sixth Marcel Grossmann Meeting on General 
Relativity}, M. Sato and T. Nakamura, eds. (World Scientific, 1992); 
D. Cangemi and R. Jackiw, \PRL{69}, 233 (1992).

\endref
\end